\documentstyle[12pt]{article}
\begin{document}
\newcommand{\be}{\begin{equation}}
\newcommand{\ee}{\end{equation}}
\def\barr{\begin{array}}
\def\earr{\end{array}}
\newcommand{\ra}{\rightarrow}
\newcommand{\mr}{{\stackrel{<}{\sim}}}
\def\bi{\bibitem}
\def\lsim{\:\raisebox{-0.5ex}{$\stackrel{\textstyle<}{\sim}$}\:}
\def\gsim{\:\raisebox{-0.5ex}{$\stackrel{\textstyle>}{\sim}$}\:}
\def\gev{\; \rm  GeV}
\def\eg{ {\it e.g.}}
\def\ie{ {\it i.e.}}
\title{
\begin{flushright}
IFT 20-96\\ [1.5ex]
{\large \bf hep-ph/9609475 } \\ [1.5ex]
\end{flushright}
LOOKING FOR A LIGHT HIGGS PARTICLE AT PRESENT AND FUTURE COLLIDERS
\footnote{Contribution to QS'96 Workshop, June 1996, Minsk, Belarus}\\ }
\author{
      MARIA KRAWCZYK  \\
{\it Institute of Theoretical Physics, University of Warsaw,
ul.Ho\.za 69} \\ 
{\it Warsaw, 00-681, Poland}}

\maketitle

\newcommand{\tb}{\tan \beta}

\begin{abstract}

Present data do not rule out the light neutral Higgs particle
$h$ or $A$ with mass below 40--50 GeV 
in the framework of the general 2HDM ("Model II"). 
The recent limits from LEP I on the parameters of the model, 
based on the Bjorken process
$Z\ra Z h$, Higgs
pair production $Z\ra A h$ and 
 the Yukawa process $Z\rightarrow f {\bar f} A$
($f= b$ quark or $\tau$ lepton) are presented.
Including  limits on
Higgs bosons masses from LEP I data additional constraints 
on the allowed value of $\tan\beta$ 
for   mass   below 2 GeV,
can be  obtained from the  existing  $(g-2)_{\mu}$ data. 
The improvement      
in the accuracy  by factor 20 in the  forthcoming $(g-2)_{\mu}$ 
 experiment  E821 may lead to  more stringent
 limits on mass of neutral Higgs boson up to  30 GeV, or even higher
if the mass difference between $h$ and $A$ is larger than $M_Z$. 
The exclusion/discovery potential 
of the gluon-gluon fusion in $ep$ collision at HERA is also discussed.
Already for a luminosity ${\cal L}_{ep}$=25 pb$^{-1}$ 
this measurement may lead to 
more stringent limits  on $\tb$ for the mass range 5-15 GeV, 
especially for the pseudoscalar case. 
In addition  the possible search for very light Higgs particle
in $\gamma \gamma$ fusion at low energy (10 GeV) LC   is described.
It  may improve bounds considerably compared to the present limits
for mass around between 1.5 and 8 GeV assuming the luminosity
10 fb$^{-1}$.

\end{abstract}

\section{Status of 2HDM.}

\subsection{Introduction.}
The mechanism of spontaneous symmetry breaking  proposed as
the source of mass for the gauge and fermion fields in the Standard 
Model (SM) leads to  a neutral scalar particle,
the minimal Higgs boson.  According to  the LEP I data,
based on the Bjorken process $e^+e^- \ra H Z^*$,
it   should be heavier than 66 GeV\cite{hi},
also
the MSSM neutral Higgs particles have  been
constrained by LEP1 data to be heavier than 
  $\sim$ 45 GeV \cite{lep,susy,hi}. The general two Higgs doublet
model (2HDM) may yet accommodate a very light ($ \lsim 45 \gev$)
 neutral scalar $h$ {\underline {or}} a pseudoscalar $A$ as long as
$M_h+M_A \gsim M_Z$~\cite{lep}.

The  minimal extension of the Standard Model is to include
a second Higgs doublet to the symmetry breaking
mechanism. In two Higgs doublet models 
the observed Higgs sector is enlarged to five scalars: two
neutral Higgs scalars (with masses $M_H$ and $M_h$ for heavier and
lighter particle, respectively), one neutral pseudoscalar
($M_A$), and a pair of charged Higgses ($M_{H^+}$ and $M_{H^-}$). 
 The
neutral Higgs scalar couplings to quarks, charged leptons and gauge 
bosons are 
modified with respect to analogous couplings in SM by factors that 
depend on additional parameters : $\tan\beta$, which is
the ratio of the vacuum expectation values of the Higgs doublets
 $v_2/v_1$,
and the mixing angle in the neutral Higgs sector $\alpha$. Further,
 new couplings appear, e.g. $Zh (H) A$ and $ZH^+ H^-$.

In this paper we will focus on the  appealing version of the models
with two doublets ("Model II") where one Higgs doublet
with vacuum expectation value $v_2$ couples only to the "up"
components 
of fermion doublets while the other one couples to the "down" 
components \cite{hunter}. 
{{In particular,  fermions couple to the pseudoscalar $A$
with a strength  proportional to $(\tan \beta)^{\pm1}$
whereas the coupling of the fermions to the scalar $h$
goes as $\pm(\sin \alpha/\cos \beta)^{\pm1}$, where the sign
$\pm$  corresponds to  isospin $\mp$1/2 components}}. 
In such model FCNC processes are absent 
and  the $\rho $ parameter retains its SM value at the tree level.
Note that in such scenario 
the large ratio $v_2/v_1 \sim m_{top}/m_b\gg 1$ is naturally 
expected.
 
The well known supersymmetric model (MSSM) belongs to this  class.
In MSSM the relations among the parameters required by the 
supersymmetry appear, leaving only two parameters free
(at the tree level) e.g. $M_A$ and $\tb$.
In general case, which we call the general 2 Higgs Doublet Model
 (2HDM), masses and parameters $\alpha$ and $\beta$ 
are not constrained by the model.
Therefore the same experimental data may lead to very distinct 
consequences  depending on which 
 version of two Higgs doublet extension of SM,
supersymmetric or nonsupersymmetric, is considered.

\subsection  {Present constraints on 2HDM  from LEP I.}
Important constraints  on  the
parameters of two Higgs doublet extensions of SM were obtained
in the precision measurements at LEP I. 
The current mass limit on 
{\underline {charged}} Higgs boson $M_{H^{\pm}}$=
44 GeV/c was obtained at LEP I \cite{sob}
from process $Z \ra H^+H^-$, 
which is  { {independent}} 
on the  parameters $\alpha$ and $\beta$. 
(Note that in
the MSSM version one expect 
$M_{H^{\pm}} > M_W$). 
For {\underline {neutral}} Higgs particles $h$ and 
$A$ there are two
main and complementary sources  of information at LEP I. One  
 is the Bjorken processes $Z \ra Z^*h $
which constrains  $g_{hZZ}^2 \sim \sin^2(\alpha-\beta)$,
for $M_h$ below 50-60 GeV..
The second  process is   $Z\ra hA$,
constraining the $g_{ZhA}^2 \sim cos^2(\alpha-\beta)$ 
for $M_h+M_A\mr M_Z$
{\footnote {
 The off shell production could also be included,
 $\eg$ as in  \cite{susy}.}}.
 This Higgs pair production contribution depends 
also on the masses $M_h$, $M_A$ and $M_Z$.

Results on $\sin^2(\alpha-\beta)$ and $\cos^2(\alpha-\beta)$   
can be translated into
the limits on neutral Higgs bosons masses $M_h$ and $M_A$.
 In the MSSM, due to relations among parameters, 
the above data allow to draw limits for the masses
 of {\underline {individual}}
 particles: $M_h\ge 45$ GeV for any $\tan \beta $ 
 and $M_A \ge$ 45 GeV for $\tan\beta \ge$1 \cite{susy,hi}.
In the general 2HDM the implications are quite different, here 
the large portion of the ($M_h$,$M_A$) plane,
where {\underline {both}} masses are in the range between 
0 and $\sim$50 GeV, is excluded \cite{lep}.

The third basic  process  in  search of a neutral  
Higgs particle at LEP I is the Yukawa process, $\ie$
 the bremsstrahlung production   of the neutral Higgs
boson $h(A)$ from the heavy fermion, 
$e^+e^- \rightarrow f {\bar f} h(A)$, where $f$ means here
{\it b} quark or $\tau$
lepton.
This process plays a very important role since 
it  constrains the production of a very light pseudoscalar
even if the pair production is forbidden kinematically,
 $\ie$ for $M_h+M_A>M_Z$ {\footnote{neglecting the off shell production}}.  
It allows also to look for a light scalar, being an additional,
and in case of $\alpha=\beta$ the most important, source of information.
 The importance  of this process was stressed in many papers\cite{pok,gle},  
 the recent  discussion of the potential of the Yukawa process
 is presented in Ref.\cite{kk}.

{{ New analysis of the Yukawa process by  
 ALEPH collaboration \cite{alef} led to 
 the exclusion  plot (95\%)   on the $\tb$ versus the  
 pseudoscalar mass, $M_A$. (Analysis by L3 collaboration 
 is also in progress { \cite{l3prep}}.). 
 It happened that obtained  limits are rather weak
{\footnote{Note, that the obtained limits
are  much weaker than the limits estimated  in Ref.
\cite{kk}.}},
 allowing for the existence of a light $A$ with mass below 10 GeV
 with $\tb$ = 20--30 , for $M_A$=40 GeV $\tb$ till 100 is allowed
!
For mass range above 10 GeV,
similar exclusion limits should in principle hold also for a 
scalar $h$ with 
the replacement in coupling $\tb\ra \sin\alpha/\cos\beta$.
Larger differences  one would expect however 
in region of lower mass,  where the production
rate at the same value of coupling 
for  the scalar is considerably larger   than 
for the pseudoscalar and therefore more 
stringent limits should be obtained \cite{kk}.

\subsection {The  2HDM  with a light Higgs particle.}

In light of the above results from precision experiments
at  LEP I
 there is still the possibility of the
 existence of one
light neutral Higgs particle with mass below $\sim$ 40--50 GeV.
As far as  other experimental data, 
especially from  low energy measurements, are concerned
they do  not  contradict 
this possibility as they 
cover only part of the parameter space of 2HDM, moreover
some of them  like the Wilczek process
 have large theoretical uncertainties
both due to  the QCD and relativistic corrections \cite{wil,hunter}
(see also discussion in \cite{bk,ames}).

In following we will study the 2HDM assuming
that one  light Higgs particle  may exist.
Moreover we will  assume  according to  LEP I data
the following mass relation between the lightest
neutral Higgs particles: $M_h+M_A \ge M_Z$.
 We specify the model further  by choosing particular 
values for the parameters $\alpha$ and $\beta$
within the present limits from
LEP I. Since $\sin(\alpha-\beta)^2$ was found \cite{lep,hi}
to be
smaller than 0.1 for the $0\mr  M_h\mr$ 50 GeV,
and  even below 0.01 for a lighter scalar, 
we simply  take  $\alpha=\beta$.
It leads to equal in  
 strengths of the coupling of  fermions to scalars   and pseudoscalars.
For the scenario with 
large $\tan\beta \sim {\cal O}(m_t/m_b)$  large
 enhancement 
in the coupling of both $h$ and $A$ bosons to the down-type
 quarks and leptons is expected.

As we described above the existing 
limits from LEP I for
a light  neutral Higgs scalar/pseudoscalar 
boson  in 2HDM are rather weak. 
Therefore it is extremely important to check if  more stringent
limits can be obtained from  other measurements.

In Sec.2 
 we present   how one can obtained 
 the limits on the parameters of the  2HDM from current  
 precision $(g-2)$ for muon  data\cite{pres},
 also
 the potential of the future E821 experiment \cite{fut}  with the accuracy
expected to be more than 20 times better is discussed.
 (See  Ref.\cite{g22} for details.)
Note that in \cite{g22}  we  took into account
the full contribution from 2HDM, i.e. exchanges of  
$h$, $A$ and $H^{\pm}$ bosons incorporating the present
constraints on Higgs bosons masses from LEP I.      
In this talk  we present limits on $\tb$ which 
can be obtained in a simple
approach (Ref.\cite{ames,deb12} and also \cite{gle}),\ie ~from the 
individual $h$ or $A$ terms. This approach reproduces  the full 2HDM
prediction up to say 30 GeV if the mass difference
between $h$ and $A$ is $\sim M_Z$, in wider range mass if 
this difference is larger. 

The possible exclusion/discovery potential of the gluon-gluon  
fusion at $ep$ collider HERA \cite{bk,ames}(Sec.3) 
and of the $\gamma \gamma$
collision at the suggested low energy LC (Sec.4)
will also be discussed {\cite{deb12}}.
In Sec.5 the combined exclusion plot (95 \% C.L.) is presented.
The search of a light neutral Higgs
particle  in heavy ion collisions at HERA and LHC are discussed 
elsewhere\cite{bol}.

\section{Constraints on  the parameters of 2HDM from $(g-2)$.}

\subsection{Present limits.}

The present experimental data limits on $(g-2)$ for muon, 
averaged over the sign of the muon electric charge, is given by \cite{data}:
$$a_{\mu}^{exp}\equiv{{(g-2)_{\mu}}\over{2}}=1~165 ~923~(8.4)\cdot 10^{-9}.$$
The quantity within parenthesis, $\sigma_{exp}$, refers to the uncertainty
in the last digit. The expected new high-precision E821 Brookhaven
experiment has design sensitivity of $\sigma_{exp}^{new}=
4\cdot 10^{-10}$ (later even 1--2 $\cdot 10^{-10}$, see Ref.\cite{czar})
 instead of the above $84\cdot 10^{-10}$.
It is of great importance to reach similar accuracy in the theoretical
analysis.

The theoretical prediction of the Standard Model for this quantity 
consists of the QED, hadronic and EW
 contribution:
$$a_{\mu}^{SM}=a_{\mu}^{QED}+a_{\mu}^{had}+a_{\mu}^{EW}.$$
The recent  SM calculations of $a_{\mu}$
are based on the  QED results from \cite{qed}, hadronic contribution
obtained in
\cite{mar,mk,jeg,wort,ll} and \cite{hayakawa} and
the EW results from \cite{czar,kuhto}. 
The uncertainties  of these
contributions differ among themselves considerably
(see below  and in  Ref.\cite{nath,czar,jeg,g22}).
The main discrepancy 
is observed  for  the hadronic contribution,
therefore  we will mainly consider case A, 
based on Refs.\cite{qed,ki,mar,mk,ll,czar},
with relatively small error in the hadronic 
part. For comparison the results for  case B (Refs.
\cite{ki,jeg,hayakawa,czar}) with the 2 times  
 larger error in the hadronic part is also displayed.
 (We adopt here  the notation from \cite{nath}.)

$$
\begin{array}{lrr}
case  &~{\rm {A~[in}}~ 10^{-9}]               &~{\rm {B~[in}}~ 10^{-9}] \\  
\hline
{\rm QED}   &~~~~~~~~1 ~165~847.06 ~(0.02)
  &~~~~~~~~~~~~~~~~~1 ~165~847.06 ~(0.02) \\
{\rm had}   & 69.70 ~(0.76) &  68.82 ~(1.54)  \\
{\rm EW}    & 1.51 ~(0.04)  &  1.51 ~(0.04)   \\
\hline
{\rm tot}   &1~165~9 1 8.27 ~(0.76)   & 1 ~165~9 1 7.39 ~(1.54)
\end{array} 
$$

\vspace{0.5cm}

The room for a new physics is given basically 
by  the difference between the experimental data and theoretical SM
prediction: $a_{\mu}^{exp}-a_{\mu}^{SM}\equiv \delta a_{\mu}$.
{\footnote {However in the calculation of
  $a_{\mu}^{EW}$ the (SM) Higgs scalar 
contribution is included(see discussion in\cite{g22}).}}
 
Below the difference $\delta a_{\mu} $ 
 for these two cases, A and B,
is presented together with 
 the error $\sigma$, obtained by adding the experimental 
and theoretical errors in quadrature:  
$$
\begin{array}{lcc}
case  &~{\rm {A ~[in}} ~10^{-9}]
	       &~~~~~~~~~~~~~~~{\rm {B ~[in}}~10^{-9}] \\  
\hline
\delta a_{\mu}(\sigma)    &4.73 (8.43) &~~~~~~~~~~~~~~5.61 (8.54)  \\
\hline
{\rm lim(95\%)} &-11.79\le\delta a_{\mu} \le 21.25      &~~~~~~~~~~~~~~-11.13\le\delta a_{\mu} \le 22.35\\
{\rm lim_{\pm}(95\%)} &-13.46\le\delta a_{\mu} \le 19.94&~~~~~~~~~~~~~~ -13.71\le\delta a_{\mu} \le 20.84      
\end{array} 
$$

\vspace{0.5cm}

One can see that at 1 $\sigma$  level the difference $\delta a_{\mu}$
can be of positive and negative negative sign.
For that  beyond SM scenarios
 in which both positive and negative 
$\delta a_{\mu}$ may appear,
 the 95\% C.L. bound can be calculated straightforward
(above denoted by $lim(95\%)$). 
For the model where  the contribution of
only {\underline {one}} sign 
is physically accessible ($\ie$ positive or negative $\delta a_{\mu}$),
 the other sign being unphysical, the 95\%C.L. limits
should be calculated in different way \cite{data}.
These limits  calculated  separately for the positive and 
for the negative contributions 
 ($lim_{\pm}$(95\%)),
lead to the shift in the
 lower  and upper
bounds by  -1.3 $\cdot 10^{-9}$ up to -2.6 $\cdot 10^{-9}$
with respect  to the standard (95\%) limits.

\subsection{Forthcoming data.} 

Since  the dominate uncertainty in $\delta a_{\mu}$ 
is due to the experimental error,
the role of the forthcoming E821 experiment is  crucial
in testing the SM or  probing  a new physics.

The future accuracy of the $(g-2)_{\mu}$ experiment is expected to be
 $\sigma^{new}_{exp}\sim0.4 \cdot 10^{-9}$ or better.
One  expects also the improvement
in the calculation of the hadronic contribution
{\footnote {The improvement in the  ongoing experiments at low energy 
in expected as well.}} such
that the total uncertainty will be basically
due to the experimental error.
Below we will assume that 
the accessible range for the beyond SM contribution, 
in particular 2HDM  with  a light scalar or pseudoscalar,
would be smaller by factor 20 as compared with the present
$lim_{\pm}$95\% bounds.
So, we consider the following option for future
measurement (in $10^{-9}$):
$$
\delta a_{\mu}^{new} =  0.24, \hspace{0.5cm} 
{\rm and}\hspace{0.5cm} 
{\rm lim_{\pm}}^{new}(95\%): -0.69\le\delta a_{\mu} \le 1.00.  $$

Assuming above bounds,
we discuss below the potential of future $(g-2)$ measurement
for the constraining the 2HDM.

\subsection{2HDM contribution to $(g-2)_{\mu}$.}

As we mentioned above the difference
 between experimental and theoretical value 
for the anomalous magnetic moment for muon
we ascribe to the 2HDM contribution, so 
we take
 $\delta a_{\mu}= a_{\mu}^{(2HDM)}$ and 
 $\delta a_{\mu}^{new} = a_{\mu}^{(2HDM)}$ for present and future 
 $(g-2)$ data, respectively.

To $  a_{\mu}^{(2HDM)}$ 
contributes a scalar $h$ ($a_{\mu}^h$), pseudoscalar $A$  ($a_{\mu}^A$)
 and the charged 
Higgs boson $H^{\pm}$ ($a_{\mu}^{\pm}$).
The relevant formulae can be found in the Appendix in Ref.\cite{g22}
Each term  $a_{\mu}^{\Lambda}$ (${\Lambda}=h,~A {\rm ~or} ~H^{\pm}$)
disappears in the limit of large mass, 
at small mass the contribution reaches its maximum (or minimum if negative)
value.
The scalar contribution $a_{\mu}^h(M_h)$ is positive whereas the
pseudoscalar  boson $a_{\mu}^A(M_A)$ gives
negative contribution, also
the  charged Higgs boson contribution is 
negative. Note that  since the mass of $H^{\pm}$ is above 44 GeV 
(LEP I limit), 
its small contribution can show up
only if the sum of $h$ and $A$ contributions is small 
(see Ref.\cite{g22} for details).

Here we present results based on a simple 
calculation of the $  a_{\mu}^{(2HDM)}$  
in two scenarios:                               
\begin{itemize}

\item {\sl a)} pseudoscalar $A$ is light, and 
$$   a_{\mu}^{(2HDM)}(M_A)= a_{\mu}^A(M_A)
~~~~~~~(1a)$$
\item {\sl b)} scalar $h$ is light,  and
 
$$   a_{\mu}^{(2HDM)}(M_h)= a_{\mu}^h(M_h)
 ~~~~~~~~(1b)$$

\end{itemize}

This simple approach is based on the 
 LEP I mass limits for charged nad neutral Higgs particles         
 and it means that $h (A)$ and $H^{\pm}$ 
are heavy enough in order to neglect 
their contributions in (1a(b)).
 The full 2HDM predictions for these two scenarios
are studied in Ref.\cite{g22}, and differences between
two approaches start to be significant above mass, say 30 GeV.

Note that the contribution 
is for the  scenario {\sl b)} positive,
 whereas for the scenario {\sl a)}~--
negative.
Therefore we have to include  this fact  
when the 95\% C.L. bounds of $  a_{\mu}^{(2HDM)}$ are calculated
(limits $lim_{\pm}(95\%)$
introduced in Sec.2.1).
Since the case A gives  more stringent $lim_{\pm}(95\%)$ constraints,
this case was used in constraining parameters of the 
2HDM.

The obtained 95\%C.L. exclusion plots for $\tb$ for 
light $h$ or  $A$  
is presented in Fig.1, together with others limits.
 The discussion of these results will be given in Sec.5.
 \begin{figure}[ht]
\vskip 4.45in\relax\noindent\hskip -1.05in
	     \relax{\includegraphics{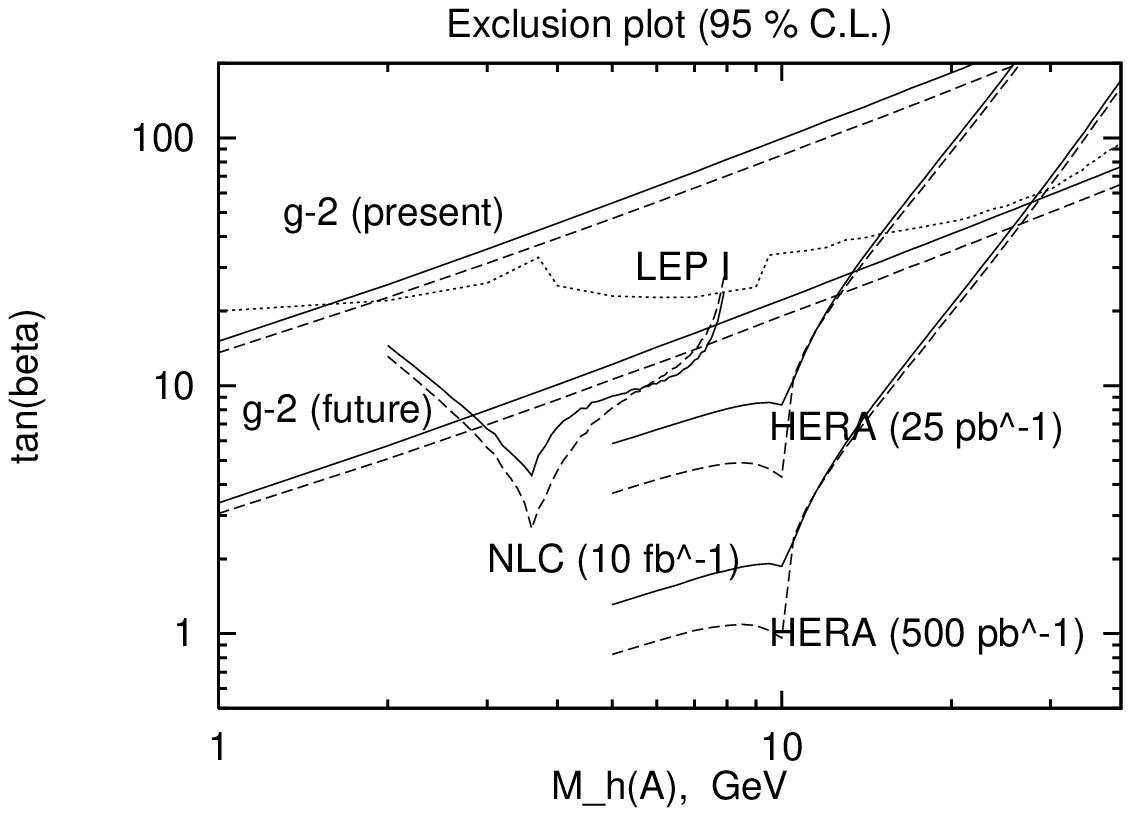}}
\vspace{-12.5ex}
\caption{ {\em The 95\% exclusion plots for  light 
scalar(solid lines) or light pseudoscalar (dashed lines)
in 2HDM. 
The limits {derivable} from present 
$(g - 2)_{\mu}$ measurement and from existing 
LEP I results (Yukawa process) for the pseudoscalar 
(dotted line) are shown. The possible exclusions  from HERA
measurement  (the gluon-gluon fusion via a quark loop with the
$\tau^+\tau^-$ final state)
for luminosity 25 pb$^{-1}$ and 500 pb$^{-1}$ as well from 
$\gamma \gamma \ra \mu^+ \mu^-$ at low energy NLC 
(10 fb$^{-1}$) are also presented.
Parameter space above the curves can be ruled out.
}}
 \label{fig:excl}
\end{figure}

\section{ Gluon-gluon fusion at HERA}

The gluon-gluon fusion
  via a quark loop,
$gg \ra h(A)$,
can be  a significant source of light non-minimal neutral Higgs bosons
at HERA  collider due to the hadronic
interaction of quasi-real photons with protons\cite{bk}.
In addition the  production of the neutral Higgs boson via
$\gamma g \ra b {\bar b} h (A)$
may also be substantial\cite{grz,bk}. Note that the latter process
also includes  
 the lowest order contributions due to the resolved photon,
like $\gamma b\ra bh(A)$, $b {\bar b}\ra h(A)$, $bg \ra h(A)b$
etc. 
We study  the potential of both $gg$ and $\gamma g$ fusions
at HERA collider. 
It was found  that for mass below $\sim 30$  GeV the $gg$
fusion via a quark loop clearly dominates the cross section.
In order to detect the Higgs particle 
it is useful to
study the rapidity distribution ${d \sigma }/{dy}$ of the Higgs bosons
in the $\gamma p$ centre of mass system.
Note that $y=-{{1}\over{2}}log{{E_h-p_h}\over{E_h+p_h}}
=-{{1}\over{2}}log{{x_{\gamma}}\over{x_p}}$, where $x_p(x_{\gamma})$
are the ratio of energy of gluon to the energy of the proton(photon),
respectively.
The (almost)
symmetric shape of  the rapidity distribution found for the signal
is extremely useful
 to reduce the
background and to separate the $gg\rightarrow h(A)$ contribution.

The main background for the Higgs mass range between
$\tau \tau$ and $b b$
 thresholds is due to $\gamma\gamma\rightarrow \tau^+\tau^-$.
In the region of negative rapidity 
the cross section ${d \sigma }/{dy}$ is very large, 
\eg ~for the $\gamma p$ energy equal to 170 GeV $\sim$ 800 pb
at the edge of phase space $y\sim -4$, then it falls down rapidly
approaching $y=0$. At the same time signal reaches at most 10 pb
(for $M_h$=5 GeV).
The region of positive rapidity  is {\underline {not}} allowed 
kinematically for this process since here one photon interacts directly with 
$x_{\gamma}=1$, and therefore $y_{\tau^+ \tau^-}
 =-{{1}\over{2}}log{{1}\over{x_p}}\leq 0$.
 Moreover, there is a
 relation between rapidity and invariant mass:
 $M^2_{\tau^+ \tau^-}=e^{2y_{\tau^+ \tau^-}}S_{\gamma p}$.
Significantly different topology found for 
 $\gamma\gamma\rightarrow \tau^+\tau^-$ events
than for the signal  
 allows to get rid of this background.
 The other sources of background are 
$q\bar q\rightarrow\tau^+\tau^-$ processes.
These processes contribute to positive and negative
rapidity $y_{\tau^+ \tau^-}$, with a flat and
relatively low cross sections in the central region (see \cite{bk}).

Assuming that the luminosity ${\cal L}_{ep}$=250 pb$^{-1}$/y 
 we predict that $gg$ fusion 
will produce approximately thousand events
per annum for $M_h=5$ GeV (of the order of 10 events  for
$M_h=30$ GeV).
A clear signature for the tagged case with $\tau^+\tau^-$ final state 
 at positive centre-of-mass
rapidities of the Higgs particle should be seen, even for the mass 
of Higgs particle above the $bb$ threshold
(more details can be found in Ref.\cite{bk}).

To show the potential of HERA collider the exclusion plot
based on the $gg$ fusion via a quark loop 
can be obtained. In this case, as we mentioned above, 
it is easy to find the
part of the phase space where the background is negligible.
To calculate the 95\% C.L. for allowed value of $\tb$
we take into account signal events corresponding only to 
the positive rapidity region (in the $\gamma p$ CM system).
Neglecting here the background   the number of events
were taken to be  equal to 3. 
 The  results
for the $ep$ luminosity ${\cal L}_{ep}$
=25 pb$^{-1}$ and 500 pb$^{-1}$
are presented in Fig. 1 and will be discused in Sec.5.

\section{Photon-photon fusion at NLC}

The possible search for a {\it very} light Higgs particle may
in principle be performed at low energy option of LC suggested
in the literature.
In the papers \cite{deb12} we addressed this problem
and find that the exclusion based on the $\gamma \gamma$ fusion
into Higgs particle decaying into $\mu \mu$ pair, 
at energy $\sqrt {s_{ee}}$=10 GeV,  may be very efficient 
in probing the value of tan $\beta$  down to 5 at $M_h\sim 3.5$ GeV
and below 15 for $2 \lsim M_h\lsim 8$ GeV 
provided that the luminosity is equal to 10 fb$^{-1}$/y (See Fig.1).

\section{Exclusion plots for 2HDM and conclusion}
In Fig.1 the  95\% C.L. exclusion curves for the $\tb$ 
in the general  2HDM ("Model II") 
obtained by us  for a light scalar (solid lines)
and for a light pseudoscalar (dashed lines)
are presented in mass range below 40 GeV.
For comparison results from LEP I analysis presented recently
by ALEPH collaboration for pseudoscalar is also shown (dotted line).
The region of ($\tb, M_{h(A)}$) above curves is excluded.

Constraints on $\tb$ 
were obtained from the  
existing  $(g-2)_{\mu}$ data including LEP I mass limits.  
We applied here a simple approach, 
which reproduces the full 2HDM contributions 
studied in Ref.\cite{g22} 
below mass of 30 GeV. 
We see  that already the present   $(g-2)_{\mu}$ data
 improve limits 
 obtained recently by ALEPH collaboration 
 on $\tb$ for low mass of the pseudoscalar:  $M_A \le$ 2 GeV.
Similar situation  should  hold for a 2HDM with a light scalar,
although here the Yukawa process may be more restrictive for $M_h\le$
10 GeV\cite{kk}.
  
The  
future improvement      
in the accuracy  by factor 20 in the  forthcoming 
$(g-2)_{\mu}$ experiment  may lead to more stringent limits 
than provided by LEP I  up to mass of a neutral Higgs boson $h$ or $A$ 
equal to 30 GeV,
if the mass difference between scalar and pseudoscalar is $\sim M_Z$,
or to  higher mass for a larger mass difference.
Note however that there is some arbitrarilness in the deriving the
expected bounds for the $\delta a_{\mu}^{new}$. 

The search at HERA in the gluon-gluon fusion via a quark loop
search at HERA may lead to even more stringent
limits (see Fig.1) for the mass range 5--15 (5--25) GeV, provided
the luminosity will reach 25 (500) pb$^{-1}$ and the 
efficiency for $\tau^+ \tau^-$ final state will be high
enough \footnote{In this analysis the 100\% efficiency
has been assumed. If the efficiency will be 10 \% the corresponding 
limits will be larger by factor 3.3}.
The other production mechanisms like the $\gamma g$ fusion
and processes with the resolved photon are expected to improve
farther these limits. 

In the very low mass range the 
additional limits can be obtained from the low energy
NL $\gamma \gamma$ collider. In Fig.1 the 
at luminosity 100 pb$^{-1}$ and 10 fb$^{-1}$.

\vspace{0.5cm}
To conclude, in the framework of 2HDM 
a light neutral Higgs scalar or pseudoscalar,
in mass range below 40 GeV,
is not ruled out by the present data. 
The future experiments may clarify the status of the 
general 2HDM with the light neutral Higgs particle. 

The role of the forthcoming g-2 measurement seems to be crucial in
clarifying which scenario of 2HDM is allowed:  with light scalar or
with light pseudoscalar.
If the $\delta a_{\mu}$ is positive/negative then the light
pseudoscalar/scalar is no more allowed.
Then  farther constraints on the coupling of the allowed light Higgs
particle one can obtained from 
the HERA collider, which   is very well suitable  for this.
The simple estimation based on one particular production mechanism
namely gluon-gluon fusion is already promising,
when adding more of them the situation may improve further\cite{bk}.
It suggests that the discovery/exclusion
potential of HERA collider is very large\cite{hera}.

The very low energy region of mass may be studied
in addition in LC machines. 
We found that the exclusion based on the $\gamma \gamma$ fusion
into Higgs particle decaying into $\mu \mu$ pair, 
at energy $\sqrt {s_{ee}}$=10 GeV,  may be very efficient 
in probing the Higgs sector of 2HDM even for luminosity 100 pb$^{-1}$.
It is not clear however if these
low energy options will come into operation.

\section{Acknowledgements}
I am grateful very much to Organizing Committee for their
kind invitation to this interesting Workshop.
The results were obtained in the collaboration with D. Choudhury
and J. \.Zochowski. Some of them  are updated according to the
reports presented during the conference ICHEP'96, July 1996, Warsaw.

\end{document}